\documentstyle[12pt]{article}

\begin{document}
{\sf \begin{center} \noindent
{\Large \bf Riemannian isometries of twisted magnetic flux tube metric and stable current-carrying solar loops}\\[3mm]

by \\[0.3cm]

{\sl L.C. Garcia de Andrade}\\

\vspace{0.5cm} Departamento de F\'{\i}sica
Te\'orica -- IF -- Universidade do Estado do Rio de Janeiro-UERJ\\[-3mm]
Rua S\~ao Francisco Xavier, 524\\[-3mm]
Cep 20550-003, Maracan\~a, Rio de Janeiro, RJ, Brasil\\[-3mm]
Electronic mail address: garcia@dft.if.uerj.br\\[-3mm]
\vspace{2cm} {\bf Abstract}
\end{center}
\paragraph*{}
Two examples of the use of differential geometry in plasma physics
are given: The first is the computation and solution of the
constraint equations obtained from the Riemann metric isometry of
the twisted flux tube. In this case a constraint between the Frenet
torsion and curvature is obtained for inhomogeneous helical magnetic
flux tube axis. In the second one, geometrical and topological
constraints on the current-carrying solar loops are obtained by
assuming that the plasma filament is stable. This is analogous to
early computations by Liley [(Plasma Physics (1964)] in the case of
hydromagnetic equilibria of magnetic surfaces. It is shown that
exists a relationship between the ratio of the current components
along and cross the plasma filament and the Frenet torsion and
curvature. The computations are performed for the helical plasma
filaments where torsion and curvature are proportional. The
constraints imposed on the electric currents by the energy stability
condition are used to solve the remaining magnetohydrodynamical
(MHD) equations which in turn allows us to compute magnetic helicity
and from them the twist and writhe topological numbers. Magnetic
energy is also computed from the solutions of MHD
equations.\vspace{0.5cm} \noindent {\bf PACS numbers:}
\hfill\parbox[t]{13.5cm}{02.40.Hw-Riemannian geometries}

\newpage
\section{Introduction}
 The tools of differential geometry and ,in particular,  Riemannian geometry \cite{1} have been proved very useful in handling
 problems in areas of physics ranging from  plasma astrophysics \cite{2} to Einstein큦 general relativity \cite{3}. This
 common feature can be better appreciated when one notices that in one of the problems we shall address in this paper, the
 isometry which is the vanishing of the variation of the metric $g_{ik}$ , or ${\delta}g_{ik}=0$ $(i,j=1,2,3)$ either in gravitational
 spacetime, or as we shall see here, in the magnetic flux tube axis geometry. Another application of differential geometry
 and topology to plasma physics is the investigation of topology and geometrical constraints in fast kinematic
 dynamos \cite{4,5} where writhe number and twist were computed from magnetic helicity, or as shall address in this paper, to the
 investigation of the stability of solar loops in plasma
 astrophysics. Riemannian geometry has also been used with great
 success to investigate the tokamaks and stellarators metric oscillations and the instabilities of confined plasmas \cite{6}.
 In this paper we  show that when one imposes the constraint of
 twisted magnetic flux tube \cite{7} metric isometry a relation
 between Frenet curvature \cite{8} and inhomogeneous helical solar loops are
 obtained. The helicoidal form of the device implies that Frenet torsion of
 magnetic field lines is compulsory. Earlier Liley큦 \cite{9} has computed
 the current conditions for static hydromagnetic equilibria. In plasmas the perturbation of energy ${\delta}W$ of magnetic surfaces
 is computed, and one is able to say if the magnetic surface is stable or unstable according the ${\delta}W\ge0$ or ${\delta}W<0$ respectively.
 The extreme condition ${\delta}W=0$ is considered to be the case of stable equilibria  in the case
understand the mechanism which allow the highly twisted coronal
magnetic flux tube to emerge from the solar surface and produce
those beautiful solar flares and loops. Recently Berger and Prior
\cite{10} have presented a detailed discussion of twist and writhe
of open and closed curves presenting a formula for the twist of
magnetic filaments in terms of parallel electric currents. s paper
we also compute the writhe and twist numbers investigated previously
by Moffat and Ricca \cite{11} and Berger and Field \cite{12} to
obtain an expression for the writhe number for the magnetically
perturbed vortex filamentary twisted structure. To obtain the writhe
number we make use of the helicity local expression for the magnetic
field proportional to the magnetic field itself and decompose the
magnetic vector field along the Serret-Frenet frame in 3D
dimensions. This allows us to solve scalar MHD equations to obtain
constraints on the global helicity expression which will allow us to
compute the writhe number. The knowledge of tilt , twist and writhe
of solar filaments for example has recently helped solar physicists
\cite{13} to work out data obtained from the vector magnetograms
placed in solar satellites. This is already an strong motivation to
go on investigating topological properties of these filamentary
twisted magnetic structures. Earlier the Yokkoh solar mission has
shown that the sigmoids which are nonplanar solar filaments are
obtained due to the action of electric currents along these
filaments which was used recently \cite{14} as motivation to
investigate current-carrying torsioned twisted magnetic curves. The
paper is organized as follows: In section 2 we compute the
conditions imposed on the current-carrying solar loops by their
stability. Section 3 addresses the computation of writhe and twist
numbers from the magnetic helicity while in section 4 we investigate
the Riemann isometry constraint on the twisted thin filament and
solve the scalar equations obtained for the curvature and torsion of
the magnetic axis of the flux tube. In section 5 we present the
conclusions.
 \section{Stability of current-carrying solar loops}
 Let us now start by considering the MHD field equations
\begin{equation}
{\nabla}.\vec{B}=0 \label{1}
\end{equation}
\begin{equation}
{\nabla}.{\vec{j}}= 0 \label{2}
\end{equation}
where ${\alpha}$ is the magnetic twist and the magnetic field
$\vec{B}$ along the filament is defined by the expression
$\vec{B}=B_{s}\vec{t}$ and
$\vec{j}=j_{s}\vec{t}+j_{n}\vec{n}+j_{b}\vec{b}$ is the electric
current density where $j_{s}$ and $j_{b}$ are respectively the
components along and cross to the solar filament all along its
extension, and $B_{s}$ is the component along the arc length s of
the filament. The vectors $\vec{t}$ and $\vec{n}$ along with
binormal vector $\vec{b}$ together form the Frenet frame which obeys
the Frenet-Serret equations
\begin{equation}
\vec{t}'=\kappa\vec{n} \label{3}
\end{equation}
\begin{equation}
\vec{n}'=-\kappa\vec{t}+ {\tau}\vec{b} \label{4}
\end{equation}
\begin{equation}
\vec{b}'=-{\tau}\vec{n} \label{5}
\end{equation}
the dash represents the ordinary derivation with respect to
coordinate s, and $\kappa(s,t)$ is the curvature of the curve where
$\kappa=R^{-1}$. Here ${\tau}$ represents the Frenet torsion. We
follow the assumption that the Frenet frame may depend on other
degrees of freedom such as that the gradient operator becomes
\begin{equation}
{\nabla}=\vec{t}\frac{\partial}{{\partial}s}+\vec{n}\frac{\partial}{{\partial}n}+\vec{b}\frac{\partial}{{\partial}b}
\label{6}
\end{equation}
 The other equations for the other legs of the Frenet frame are
\begin{equation}
\frac{\partial}{{\partial}n}\vec{t}={\theta}_{ns}\vec{n}+[{\Omega}_{b}+{\tau}]\vec{b}
\label{7}
\end{equation}
\begin{equation}
\frac{\partial}{{\partial}n}\vec{n}=-{\theta}_{ns}\vec{t}-
(div\vec{b})\vec{b} \label{8}
\end{equation}
\begin{equation}
\frac{\partial}{{\partial}n}\vec{b}=
-[{\Omega}_{b}+{\tau}]\vec{t}-(div{\vec{b}})\vec{n}\label{9}
\end{equation}
\begin{equation}
\frac{\partial}{{\partial}b}\vec{t}={\theta}_{bs}\vec{b}-[{\Omega}_{n}+{\tau}]\vec{n}
\label{10}
\end{equation}
\begin{equation}
\frac{\partial}{{\partial}b}\vec{n}=[{\Omega}_{n}+{\tau}]\vec{t}-
\kappa+(div\vec{n})\vec{b} \label{11}
\end{equation}
\begin{equation}
\frac{\partial}{{\partial}b}\vec{b}=
-{\theta}_{bs}\vec{t}-[\kappa+(div{\vec{n}})]\vec{n}\label{12}
\end{equation}
Now that we have this mathematical machinery at our disposal, let us
consider the Bernstein et al \cite{15} plasma energy stability
relation
\begin{equation}
{\delta}W=\frac{1}{2}\int{ds[(\vec{Q}+\vec{n}.\vec{\epsilon}\vec{j}{\times}\vec{n})^{2}+{\gamma}p(div{\vec{\epsilon}})^{2}-2(\vec{n}.\vec{\epsilon})^{2}\vec{j}{\times}\vec{n}.\vec{B}.{\nabla}\vec{n})]}\label{13}
\end{equation}
where
\begin{equation}
\vec{Q}=curl(\vec{\epsilon}{\times}\vec{B}) \label{14}
\end{equation}
and ${\gamma}$ is the ratio of the specific heats. According to
Liley큦 \cite{14} if one puts
\begin{equation}
S= \vec{j}{\times}\vec{n}.(\vec{B}.{\nabla}\vec{n})\label{15}
\end{equation}
and if there if there is no discontinuities in the equilibrium
values, then a sufficient condition for stability , i.e,
${\delta}W{\ge}0$, is
\begin{equation}
S{\le}0 \label{16}
\end{equation}
substitution of the expressions for $\vec{j}$ and for $\vec{B}$
\begin{equation}
\vec{j}{\times}\vec{n}=j_{b}\vec{t}-j_{s}\vec{b}\label{17}
\end{equation}
into this expression yields the following geometrical constraint in
the solar loop
\begin{equation}
[j_{b}\kappa(s)+\tau(s)j_{s}]B\ge0\label{18}
\end{equation}
By assuming that $B\ge0$ one obtains in the stable equilibrium case
${\delta}W=0$ or $S=0$ which reads
\begin{equation}
\frac{j_{b}}{j_{s}}=-\frac{\tau}{\kappa}\label{19}
\end{equation}
Note that in planar (torsionless) solar loops the normal current
$j_{b}$ vanishes. In helical solar loops where the ratio between
torsion and curvature is constant or ${\beta}=-\frac{\kappa}{\tau}$
one has the following relation
\begin{equation}
\frac{dj_{s}}{ds}=-{\beta}\frac{dj_{b}}{ds}\label{20}
\end{equation}
Physically  this means that as observe the solar loop along its
extension the toroidal current increases while the normal current
decreases as going to a focal point. A long and straightforward
computation allows us to write down the following expression
\begin{equation}
{\nabla}.\vec{j}=0\label{21}
\end{equation}
implies
\begin{equation}
{\partial}_{s}{j}_{s}+[{\theta}_{bs}+div\vec{b}]{j}_{s}=-j_{b}(div\vec{b}+{\theta}_{bs})
\label{22}
\end{equation}
substitution of expression (\ref{19}) into (\ref{22}) allows us to
obtain a differential equation for the current $j_{s}$ as
\begin{equation}
{\partial}_{s}{j}_{s}+{\theta}_{bs}[1-\frac{{\tau}_{0}}{{\kappa}_{0}}]{j}_{s}=0
\label{23}
\end{equation}
which yields the solution
\begin{equation}
{j}_{s}={j_{s}}^{0}exp[-\int{{\theta}_{bs}(1-\frac{{\tau}_{0}}{{\kappa}_{0}})ds}]
\label{24}
\end{equation}
Note that the current along the solar loop decays as the coordinate
$s{\rightarrow}\infty$.  From  this last expression one notes that
\begin{equation}
\frac{{j}_{s}}{{j_{s}}^{0}}=\frac{exp[-\int{{\theta}_{bs}ds}]}{exp[-\int{\frac{{\tau}_{0}}{{\kappa}_{0}}ds}]}\le1
\label{25}
\end{equation}
where the RHS inequality is valid since the initial current is
greater than the resultant current. This inequality is equivalent to
\begin{equation}
{exp[-\int{{\theta}_{bs}ds}]}\le{exp[-\int{\frac{{\tau}_{0}}{{\kappa}_{0}}ds}]}
\label{26}
\end{equation}
which yields the following constraint between the ratio between
torsion and curvature and the factor ${\theta}_{bs}$ as
\begin{equation}
{\theta}_{bs}\ge\frac{{\tau}_{0}}{{\kappa}_{0}} \label{27}
\end{equation}
in all points of the solar loops which are possibly broken in the
inflexionary points of the loops \cite{7}. our computations we
neglect $div{\vec{b}}$ and also consider that ${\theta}_{bs}$ varies
very slowly along the solar loop. Now let us compute the magnetic
field from the equation ${\nabla}.\vec{B}=0$ which becomes
\begin{equation}
{\partial}_{s}{B}_{s}+[{\theta}_{bs}+{\theta}_{ns}]{B}_{s}=0
\label{28}
\end{equation}
A simple solution of this equation can be obtained as
\begin{equation}
{B}_{s}=B_{0}exp[\int{({\theta}_{bs}+{\theta}_{ns})ds}] \label{29}
\end{equation}
where $B_{0}$ is an integration constant. Substitution into equation
(\ref{17}) assuming that the flow is geodesic along the filament and
abnormality relation ${\Omega}_{s}=0$ along with
${\theta}_{bs}={\theta}_{ns}=0$one obtains
\begin{equation}
{B}_{n}=B_{0}[(div{\vec{b}})^{-1}-\int{{\kappa}ds}] \label{30}
\end{equation}
In the next section we shall compute the magnetic energy in terms of
the magnetic vector potential $\vec{A}$, which allows us to compute
the helicity integral and the writhe number and twist of the solar
loops.
\section{Magnetic energy, Twist and Writhe of Solar Loops}
In this section we compute the magnetic energy , twist and writhe of
the solar filament based on the vector potential, assuming that it
obeys the Coulomb gauge ${\nabla}.\vec{A}=0$ and the definition
$\vec{B}={\nabla}{\times}\vec{A}$. The Coulomb gauge becomes
\begin{equation}
{\partial}_{s}A_{s}+[{\theta}_{bs}+{\theta}_{ns}]A_{s}=0 \label{31}
\end{equation}
assuming that $A=A(s,n)$ together with the equations for the
definition of $\vec{B}$, taking $\vec{A}=A_{s,n}\vec{t}$ yields
\begin{equation}
{\partial}_{n}{A}+{\kappa}A=0 \label{32}
\end{equation}
\begin{equation}
[2\tau+{\Omega}_{n}+{\Omega}_{s}]A=0\label{33}
\end{equation}
which upon separation of variables $A={\Phi}(s){\Theta}(n)$ one is
able to find
\begin{equation}
{A(s,n)}=A_{0}exp[-{{\kappa}_{0}}n+\int{({\theta}_{bs}+{\theta}_{ns})ds}]
\label{34}
\end{equation}
Let us now compute the magnetic energy
\begin{equation}
E_{B}=\frac{1}{8{\pi}}\int{B^{2}dV} \label{35}
\end{equation}
which yields
\begin{equation}
E_{B}=\frac{{(B_{0}a)}^{2}}{8}exp[-2({\theta}_{bs}+{\theta}_{ns})s]
\label{36}
\end{equation}
Note that from above constraints
${\theta}_{bs}=\frac{{\tau}_{0}}{{\kappa}_{0}}$ and allowing
${\theta}_{ns}$ to vanish one obtains the energy
\begin{equation}
E_{B}=\frac{{(B_{0}a)}^{2}}{8}exp[-2(\frac{{\tau}_{0}}{{\kappa}_{0}})s]
\label{37}\end{equation} which is the energy of solar loops based on
the assumption of its stability. Now let us compute the global
helicity
\begin{equation}
H=\int{{\vec{A}}.\vec{B}dV} \label{38}
\end{equation}
which is equivalent to
\begin{equation}
H=\int{[A(s,n).B_{s}]dV}\label{39}
\end{equation}
substitution of the relations above yields
\begin{equation}
H=-{\pi}a^{2}\int{[2{\tau}+{\Omega}_{s}+{\Omega}_{n}]A^{2}ds}
\label{40}
\end{equation}
Besides if one considers that the plasma flow in solar loops is
geodesic the abnormality factor ${\Omega}_{s}$ vanishes and the
magnetic helicity reduces to
\begin{equation}
H=-{\pi}a^{2}\int{[2{\tau}+{\Omega}_{n}]A^{2}ds} \label{41}
\end{equation}
If one assumes that the magnetic vector potential varies also quite
slow along the solar filament one has
\begin{equation}
H=-{\pi}a^{2}A^{2}\int{[2{\tau}+{\Omega}_{n}]ds} \label{42}
\end{equation}
The first integral on the RHS in (\ref{42}) is the total torsion and
considering that the helicity is proportional to the sum of twist
and writhe and yet that the twist is proportional to the total
torsion , we conclude that the writhe number Wr is proportional to
the integral of the abnormality ${\Omega}_{n}$.
\section{Riemann isometry of magnetic flux tube metric}
Since in the limit of thin magnetic flux tube one could represent it
as a mathematical model for solar loop, it is interesting to argue
if there is a similar way to constraint the magnetic flux tube as
was done for the current-carrying solar loops in previous sections.
A simple way to address and respond to this question is to resource
to a simple tool of Riemannian geometry \cite{7,16} used in general
relativity, which is called metric isometry and is represented by
the expression
\begin{equation}
{\delta}g_{ik}=0 \label{43}
\end {equation}
If one applies the isometry hypothesis to the flux tube metric
\begin{equation}
dl^{2}=dr^{2}+r^{2}d{\theta}_{R}+{K^{2}(s)}ds^{2} \label{44}
\end{equation}
one obtains that ${\delta}g_{rr}=0$ and that
\begin{equation}
{\delta}g_{{\theta}{\theta}}=[\frac{\partial}{{\partial}r}g_{rr}]{\delta}r=2r{\delta}r=0
\label{45}
\end {equation}
which yields $r=0$ and means we are close to the magnetic axis of
the tube, while
\begin{equation}
{\delta}g_{ss}=[\frac{\partial}{{\partial}r}g_{ss}]{\delta}r+[\frac{\partial}{{\partial}s}g_{ss}]{\delta}s
 \label{46}
\end {equation}
which yields
\begin{equation}
{\delta}g_{ss}=[r(\kappa(s)sin{\theta}\frac{\partial}{{\partial}s}{\theta}-\frac{\partial}{{\partial}s}\kappa(cos{\theta}))]{\delta}s
\label{47}
\end {equation}
where we assume that ${\delta}r=0$ or that we are over the magnetic
axis of the flux tube. Here we also consider as in Ricca큦 paper
\cite{8} that ${\theta}={\theta}_{R}-\int{{\tau}ds}$. Applying the
isometry condition above for this component one finally obtains the
constraint equation for the Frenet curvature as
\begin{equation}
\frac{\partial}{{\partial}s}\kappa=-{{\tau}_{0}}^{2}s+c_{0}
\label{48}
\end {equation}
which after a simple integration yields
\begin{equation}
\frac{\partial}{{\partial}s}\kappa=-\frac{1}{2}{{\tau}_{0}}^{2}s^{2}+c_{0}s+c_{1}
\label{49}
\end {equation}
This equation was obtained under the assumption that the magnetic
axis of the twisted magnetic flux tube torsion is constant
${\tau}={\tau}_{0}$ and that $c_{0}$ and $c_{1}$ are integration
constants.

\section{Conclusions}
 In conclusion, topological and geometrical constraints on solar loops were obtained through the use of the tools of
 differential geometry. Other kinds of isometries such as the
 Riemann curvature tensor $R_{ijkl}$ isometry ${\delta}R_{ijkl}=0$
 could be calculated for tokamaks and stellarators as well as for
 solar loops. This may appear elsewhere. We also hope that the
 analytical models discussed here may be useful to solar physicists
 in the checking their observational results in magnetic twist and
 writhe in solar activity regions.

 \section*{Acknowledgements}
 Thanks are due to CNPq and UERJ for financial supports.

\end{document}